# Attack vulnerability of complex networks

Petter Holme* and Beom Jun Kim†
*Department of Theoretical Physics, Umeå University, 901 87 Umeå, Sweden*

Chang No Yoon and Seung Kee Han
*Department of Physics, Chungbuk National University, Cheongju, Chungbuk 361-763, Korea*

We study the response of complex networks subject to attacks on vertices and edges. Several existing complex network models as well as real-world networks of scientific collaborations and Internet traffic are numerically investigated, and the network performance is quantitatively measured by the average inverse geodesic length and the size of the largest connected subgraph. For each case of attacks on vertices and edges, four different attacking strategies are used: removals by the descending order of the degree and the betweenness centrality, calculated for either the initial network or the current network during the removal procedure. It is found that the removals by the recalculated degrees and betweenness centralities are often more harmful than the attack strategies based on the initial network, suggesting that the network structure changes as important vertices or edges are removed. Furthermore, the correlation between the betweenness centrality and the degree in complex networks is studied.

## I. INTRODUCTION

Examples of complex networks are abundant in many disciplines of science and have recently received much attention [1, 2]. Many works have tried to regenerate geometrical statistics of real-world networks by generative algorithms that mimics behaviors found in the real-world networks. The studies along this line have been able to model, e.g., the emergence of scale-free degree distributions [3] and the high clustering of social networks [4, 5]. Another group of complex network studies aims to investigate certain dynamical problems on network topologies [5, 6]. A third group of works studies how the geometric characteristics and performances of the networks are affected by the restrictions imposed on networks. The approach taken by the present paper belongs to the third category as we study the robustness of the network subject to various attack strategies.

Originated from studies of computer networks, "attack vulnerability" [3] denotes the decrease of network performance due to a selected removal of vertices or edges. In the present study we measure the attack vulnerability of various complex network models and real world networks. We compare different ways of attacking the network and use various ways of measuring the resulting damage. In general, this gives a measure of the decrease of network functionality under a sinister attack. The meaningful purpose for attack vulnerability studies is for the sake of protection: If one wants to protect the network by guarding or by a temporary isolation of some vertices (edges), the most important vertices (edges), breaking of which makes the whole network malfunctioning, should be identified. Furthermore, one can learn how to build attack-robust networks, and also how to increase the robustness of vital biological networks. Also in a large network of a criminal organization, the whole network can be made to collapse by arresting key persons, which can be identified by a similar study. However, the applicability to social networks may not be very high—acquaintance ties are to some extent subjective and time dependent [7], and when a social network structure is under attack, the dynamics would probably speed up as the organization tries to protect itself.

A topic closely related to attack vulnerability is that of the percolation on complex networks [8], where all vertices (or edges) have the equal probability of being disabled. In the network of computers this situation corresponds to a random breakdown of computers, while in the problem of the disease spread through network of people it corresponds to that a randomly chosen set of people are susceptible. One of the key quantities in percolation studies, the size of the largest connected subgraph, is also used in the present paper as one of the measures of network performance.

This paper is organized as follows: In Sec. II we provide the definitions of terms and measured quantities. In Sec. III various attack strategies are explained. In Sec. IV two real-world networks and several complex network models used in the present paper are briefly described. Sections V, VI, and VII are devoted to the main results, on the relation between degrees and betweenness centralities, on the vulnerability under vertex attack, and on the vulnerability under edge attack. Finally, we summarize our results in Sec. VIII.

*Electronic address: holme@tp.umu.se
†Electronic address: kim@tp.umu.se

## II. DEFINITIONS OF QUANTITIES

In general, the complex networks—networks of both randomness and structure—studied in this article can be represented by an undirected and unweighted graph: $\mathcal{G} = (\mathcal{V}, \mathcal{E})$, where $\mathcal{V}$ is the set of vertices (or nodes), and $\mathcal{E}$ is the set of edges (or links). Each edge connects exactly one pair of vertices, and a vertex-pair can be connected by maximally one edge, i.e., multiconnection is not allowed. Let furthermore $N$ denote the number of vertices $N = |\mathcal{V}|$ and $L$ the number of edges $L = |\mathcal{E}|$. For a social network [9], $\mathcal{V}$ is a set of persons (or 'actors' in sociology parlance) and $\mathcal{E}$ is the set of acquaintance ties that links the persons together. In computer networks $\mathcal{V}$ represent the routers or computers and $\mathcal{E}$ the channels for computer communication.

There are several ways of measuring the functionality of networks. One key quantity is the average geodesic length $\ell$, which is sometimes termed "the characteristic path length," defined by:

$$\ell \equiv \langle d(v,w) \rangle \equiv \frac{1}{N(N-1)} \sum_{v \in \mathcal{V}} \sum_{w \neq v \in \mathcal{V}} d(v,w),$$

where $d(v,w)$ is the length of the geodesic between $v$ and $w$ ($v, w \in \mathcal{V}$), i.e., the number of edges in the shortest path connecting the two, and the factor $1/N(N-1)$ is one over the number of pairs of vertices. If $\ell$ is large the dynamics (of epidemics, information flow, etc.) is slow in the network. Social networks are known to have very short average geodesic length, $\ell \propto \log N$, with the "six degrees of separation", $\ell \approx 6$, of earth's population as a celebrated example [10]. The logarithmic increase of $\ell$ is also characteristic of computer networks, and $\ell \approx 17$ has been estimated for the entire World-Wide Web [3]. As the number of removed vertices or edges is increased, the network will eventually break into disconnected subgraphs. The average geodesic length, by definition, becomes infinite for such a disconnected graph, and one can instead study the average inverse geodesic length:

$$\ell^{-1} \equiv \left\langle \frac{1}{d(v,w)} \right\rangle \equiv \frac{1}{N(N-1)} \sum_{v \in \mathcal{V}} \sum_{w \neq v \in \mathcal{V}} \frac{1}{d(v,w)}, \quad (1)$$

which has a finite value even for a disconnected graph since $1/d(v,w) = 0$ if no path connects $v$ and $w$. It should be noted that the notation $\ell^{-1}$ does not mean the reciprocal of $\ell$. The functionality of the network is then measured by $\ell^{-1}$: the larger $\ell^{-1}$ is the better the network functions.

Since subsequent attacks will disintegrate the network, the size of the largest connected subgraph is also an interesting quantity for measuring the functionality of the networks. In social networks, the largest connected subgraph is known to have a size of the order of the entire network, and accordingly is called "giant component" [11]. Throughout the present paper, we denote the size of the giant component as $S$, which will be used together with $\ell^{-1}$ to study the attack vulnerability.

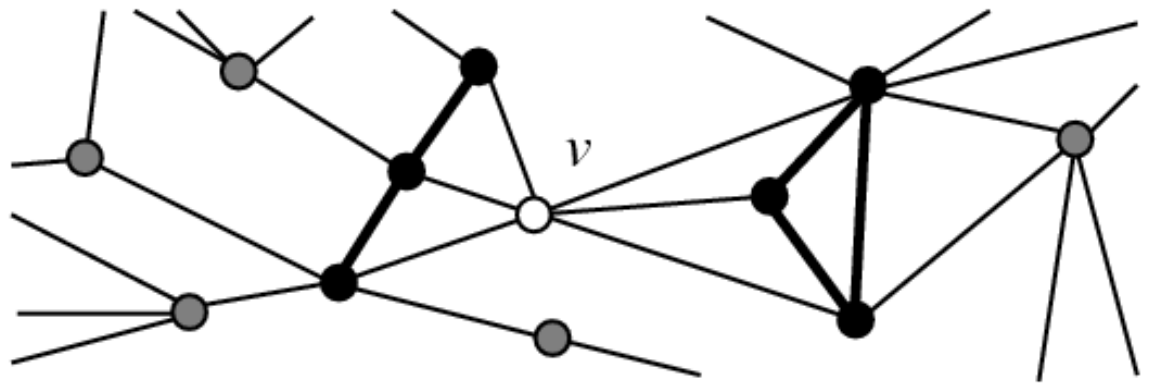

FIG. 1: An example of how to calculate the local clustering coefficient $\gamma_v$ of the vertex $v$. The full-filled black circles indicate the neighborhood $\Gamma_v$ of $v$, and the thick lines are the edges connecting two vertices within $\Gamma_v$. Since there are five such edges ($|\Gamma_v|_{\mathcal{E}} = 5$) and the degree $k_v = 6$, we obtain $\gamma_v = 1/3$ from Eq. (2).

In addition to the logarithmically increasing average geodesic length, social networks are found to have a high local transitivity: if $\{u, v\}$ and $\{u, w\}$ are two connected pairs, then $v$ is likely to be connected to $w$ too (and if it does $\{u, v, w\}$ is called a "triad") [12]. The clustering coefficient $\gamma$ (introduced in Ref. 5) intends to measure the average degree of the local transitivity in a network: Let $|\Gamma_v|_{\mathcal{E}}$ denote the number of edges in the neighborhood $\Gamma_v$ of $v \in \mathcal{V}$ then

$$\gamma_v \equiv \frac{|\Gamma_v|_{\mathcal{E}}}{\binom{k_v}{2}} \quad (2)$$

is called the local clustering coefficient of the vertex $v$. Here the degree $k_v$ of $v$ is defined as the number of vertices in $\Gamma_v$, i.e., $k_v \equiv |\Gamma_v|$. The "clustering coefficient" is then defined as the average of $\gamma_v$:

$$\gamma \equiv \langle \gamma_v \rangle \equiv \frac{1}{N} \sum_{v \in \mathcal{V}} \gamma_v. \quad (3)$$

An alternative interpretation is that $\gamma_v$ is the fraction of number of triads divided by the maximal number of triads. In Fig. 1, we present an illustration to explain the meaning of the local clustering coefficient: The number of edges within the neighborhood $|\Gamma_v|_{\mathcal{E}} = 5$ and the degree $k_v = 6$ result in $\gamma_v = 1/3$ in Fig. 1. Both $\gamma_v$ and $\gamma$ are strictly in the interval $[0, 1]$, with $\gamma = 1$ attained only for a fully connected network, where every vertex is connected to every other vertex with the total number of edges $L = N(N-1)/2$.

Removals of important vertices may affect the network significantly. For example, in Ref. 3 only a few removals of vertices with the highest degrees has been shown to be enough to alter the behaviors of scale-free networks and the average geodesic length has been found to increase dramatically. In the studies of social networks, the centrality is an important concept that tries to capture the prominence of a person in the embedding social structure. It is natural to expect that removals of vertices with high centrality will worsen the functionality of networks more than the removals by degrees. It should be noted that the vertex with a low degree can have a high centrality (this will be shown explicitly in Sec. V) and thus attacking the network by removing vertices with high centralities may differ from that by degrees. Among many centrality measures [14] we focus on the "vertex betweenness centrality" $C_B(v)$ [15] defined for a vertex $v \in \mathcal{V}$ as follows:

$$C_B(v) = \sum_{w \neq w' \in \mathcal{V}} \frac{\sigma_{ww'}(v)}{\sigma_{ww'}}, \quad (4)$$

where $\sigma_{ww'}$ is the number of geodesics between $w$ and $w'$, and $\sigma_{ww'}(v)$ is the number of geodesics between $w$ and $w'$ that passes $v$. Similarly, one can define the "edge betweenness centrality" $C_B(e)$ for an edge $e \in \mathcal{E}$ as

$$C_B(e) = \sum_{w \neq w' \in \mathcal{V}} \frac{\sigma_{ww'}(e)}{\sigma_{ww'}}, \quad (5)$$

where $\sigma_{ww'}(e)$ is the number of geodesics between $w$ and $w'$ that includes the edge $e$. Throughout the present paper, we call $C_B(v)$ and $C_B(e)$ as the vertex betweenness and the edge betweenness for brevity. For calculations of $C_B(v)$ and $C_B(e)$ we use the $O(NL)$ algorithm presented in Refs. 16 and 17.

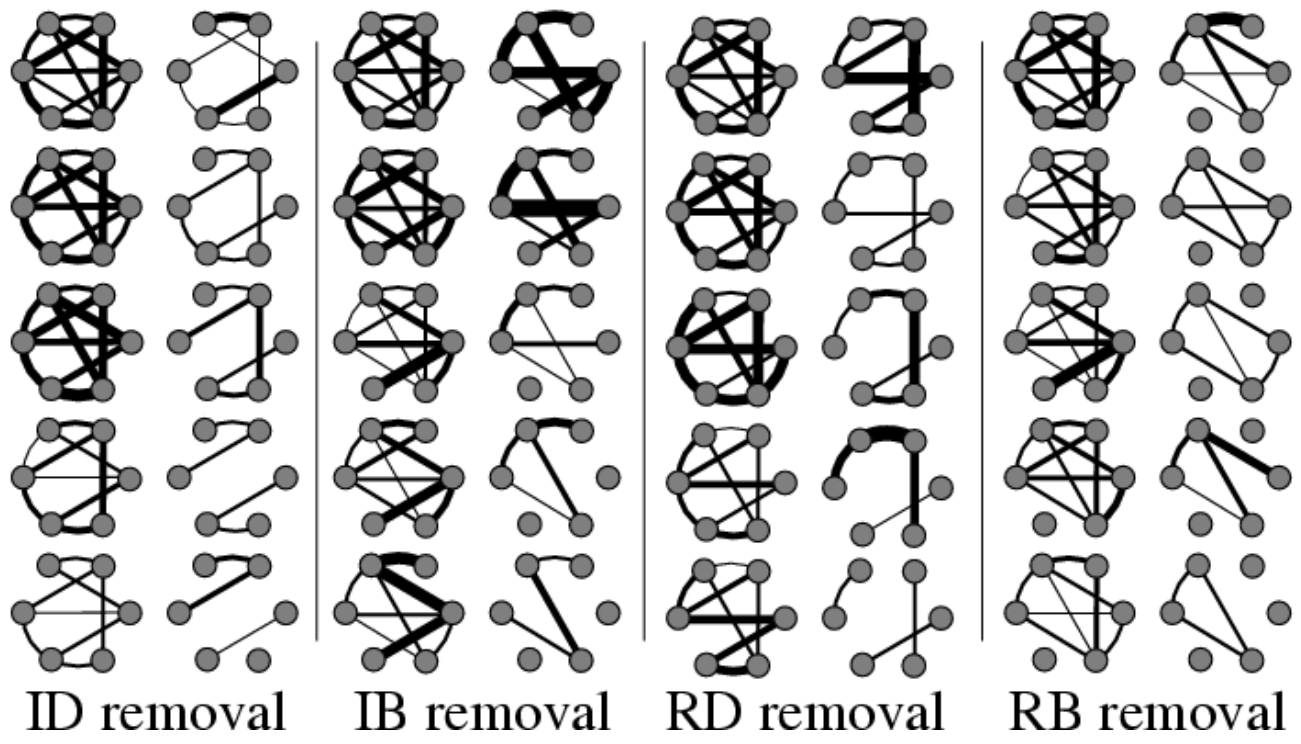

FIG. 2: Various attack strategies for edge removals applied to a realization of the generative algorithm of the Watts-Strogatz model of small-world network (see Sec. IV D). From left to right, the evolutions of network structures are shown for the edge attack strategies based on ID (initial degree), IB (initial betweenness), RD (recalculated degree), and RB (recalculated betweenness), respectively. The initial network structure is displayed at the top left corner of each column and the subsequent structures at next nine steps are exhibited (first four steps from top to bottom and then five more steps from top right to bottom right). For an individual subgraph the thickness of the lines is proportional to the betweenness of the corresponfing edge.

## III. ATTACK STRATEGIES

For the study of attack vulnerability of the network, the selection procedure of the order in which vertices are removed is an open choice. One may of course maximize the destructive effect at any fixed number of removed vertices (or edges). However, this requires the knowledge of the whole network structure and pinpointing the vertex to attack in this way makes a very time-demanding computation. A more tractable choice, used in the original study of computer networks, is to select the vertices in the descending order of degrees in the initial network and then to remove vertices one by one starting from the vertex with the highest degree [3]; this attack strategy uses the initial degree distribution and thus is called "ID removal" throughout the current paper. The vertices with high betweenness also play important roles in connecting vertices in the network. [31] The second attack strategy is called "IB removal" and uses the initial distribution of the betweenness. Both ID removal and IB removal use the information on the initial network. As more vertices are removed, the network structure changes, leading to the different distributions of the degree and the betweenness than the initial ones. The third attack strategy called "RD removal" uses the recalculated degree distribution at every removal step, and the fourth strategy, we call it "RB removal", is based on the recalculated betweenness at every step. RD removal has been used in Refs. 18, 19. It should be noted that ID and RD removals are local strategies, while the other two based on the betweenness are global ones, which makes the applications of ID and RD $O(N)$ algorithms while IB and RB are $O(NL)$ even with the best known algorithm [16, 17]. The other important difference between the degree-based and the betweenness-based strategies is that the former concentrate on reducing the total number of edges in the network as fast as possible whereas the latter concentrate on destroying as many geodesics as possible. It is not entirely clear *a priori* which one of these four different attack strategies should be more harmful than the others, although one can naively expect that RD and RB are more harmful than ID and IB, respectively. It should also be noted that for the strategies based on the recalculated information the most harmful sequences for removals of $N_{\rm rm}$ vertices and $N'_{\rm rm}$ vertices ($N_{\rm rm} \neq N'_{\rm rm}$) might differ significantly even in the early stage of attacks.

One can also attack edges instead of vertices. In the network of computers attacking edges may correspond to the cutting off the communication cables, while the attacks on vertices can be interpreted as breakdowns of servers by malicious hackers. (The opposite is of course also imaginable: a software obstruction of a communication link, or a server destroyed physically.) The vulnerability of networks under edge attacks is also studied by using similar strategies (we again call them as ID, IB, RD, and RB removals of edges). The concept of the edge betweenness was introduced in Sec. II from a straightforward generalization of the vertex betweenness. On the other hand, the definition of the "edge degree" is not so clear. But still it is expected that the importance of an edge should be possible to assess by the degrees of the two vertices it connects. In this work we attempt to define the edge degree $k_e$ from the local information of the vertex degrees in several different ways:

$$k_e \equiv k_v k_w, \quad (6a)$$
$$k_e \equiv k_v + k_w, \quad (6b)$$
$$k_e \equiv \min(k_v, k_w), \quad (6c)$$
$$k_e \equiv \max(k_v, k_w), \quad (6d)$$

where the edge $e$ connects two vertices $v$ and $w$ with vertex degrees $k_v$ and $k_w$, respectively. As will be discussed in Sec. V, among the above definitions, we find that Eq. (6a) gives a more reasonable result (a higher $C_B(e)$ to $k_e$ correlation) than the others, and thus the "edge degree" defined as $k_e \equiv k_v k_w$ is used for the attack strategies ID and RD edge removals.

From the definitions, we expect that a vertex with higher degree usually should have higher betweenness in most real-world networks. However, the correlation between edge degree and edge betweenness is less obvious. This is expected to show a larger difference between degree-based and betweenness-based attack strategies for edge attacks than for vertex attack. The four different edge attack strategies applied to a network generated by Watts and Strogatz' small-world network model (see Sec. IV D) is shown in Fig. 2. Quite soon the original network structure is lost and procedure-specific structures emerge. For example, the RB edge removal concentrates on edges of high betweenness, and thus edges which carry more geodesics are first lost. Consequently, it is not a surprise that the resulting network structure by RB consists of highly connected clusters and vertices with no neighbors. The RD procedure, on the other hand, removes edges connecting vertices with high vertex degrees, and therefore it is natural that the original network is split into many subgraphs of vertices with low (but not zero) degrees.

In Secs. VI and VII, we investigate various networks subject to the above mentioned four different attack strategies, ID, IB, RD, and RB, applied for vertex removals and edge removals. To detect the damages caused by those attacks, we measure the average inverse geodesic length $\ell^{-1}$ in Eq. (1) as well as the size $S$ of the giant component. As the vertex attack proceeds, both the remaining number of vertices and the average inverse geodesic length decrease, which, from the definition of $\ell^{-1}$, suggests that $\ell^{-1}$ can be both increasing and decreasing, depending on how much damage is made by the removals. However, the edge removals do not change the number of vertices in the network and thus $\ell^{-1}$ should be a decreasing function of the number of removed edges. Similarly, $S$ is expected to show different behaviors for vertex and edge removals. For vertex removals, $S$ versus the number of removed vertices should have a slope unity in the initial attack stages since the removed vertex probably belonged to the giant component. On the other hand, the initial edge attacks should not change the size of the giant component, and thus $S$ versus the number of removed edges should start as a horizontal line.

We conclude the section with some technical details: In any case where two or more vertices (edges) could equally be chosen by some strategy, the selection is done randomly. For the RB strategy, if the betweenness is zero for all vertices, i.e., the vertices are either isolated or linked to exactly one neighbor, the vertices with $k_v = 1$ are attacked before the meaningless attack of vertices with $k_v = 0$.

## IV. NETWORKS

To study the emergence of different geometrical properties of complex networks such as social networks, power grids, metabolic networks, computer networks and so on, different generative algorithms have been proposed [1]. Among various existing models for generating networks similar to real ones, two generic models, the Watts-Strogatz (WS) model of the small-world networks [5] and the Barabási-Albert (BA) model of the scale-free network [3, 20], have been widely studied. Both models commonly show the behavior that two arbitrarily chosen vertices are connected by a remarkably short path. More specifically, the average geodesic length has been found to scale logarithmically with the network size. On the one hand, the WS model does not exhibit the power-law distribution of degree which many real networks show and the BA model successfully produces. On the other hand, the WS model has high clustering like, e.g., social networks, whereas the BA model has a clustering coefficient that scales toward zero as $N \to \infty$. There have been attempts to revise and extend those representative models in order to produce a network model which can show the small average geodesic length, the scale-free degree distribution, and high clustering, all at the same time [4, 22]. In this work, we study two real networks, a "social network" constructed from scientific collaboration data (Sec. IV A), and a "computer network" constructed from computer traffic over the Internet (Sec. IV B), as well as four model networks, the random network model by Erdös and Rényi (ER), the WS model, the BA model, and the clustered scale-free network (CSF) model suggested by two of the present authors in Ref. 4 (Secs. IV C through IV F). It should be noted that network models such as those mentioned above, model the emergence of structure in networks—structure that can be monitored by certain quantities such as degree distribution, clustering coefficient, an so on. However, they do (probably) not sample the ensemble of networks defined by specific values of these quantities uniformly. This is known to be the case for other ways of generating random graphs with structural biases that are easier to give a probability-theoretical analysis [23]. That the sampling is biased make inference from the scaling of quantities difficult—if, e.g., one model gives networks with the same values as another except, say, a smaller clustering coefficient, it is not certain that a different behavior under, say, edge removal is due to the lower clustering.

### A. Scientific collaboration network from the hep-lat e-print archive

To obtain a well-defined social network from real world data we follow Ref. 17 and construct a network of scientific collaborations from the the Los Alamos preprint archives [24] in the following way: If two scientists wrote an article together, they are connected by an edge. Accordingly, the vertices in the network are scientists, and the edges represent the collaboration ties. For the attack vulnerability calculations, the whole Los Alamos e-print archive is too big to be computationally tractable, [32] instead we chose `hep-lat` database, which contains preprints about lattice studies in high energy physics, among various subcategories only for a computational convenience. The network used in analysis has $N = 2010$ (the number of vertices) and $L = 6614$ (the number of edges), and the size of the giant component is $S = 1412$ and the clustering coefficient is $\gamma \approx 0.571$. A discussion on the usefulness of collaboration networks as real-world social networks can be found in Ref. 17.

### B. Computer network from Internet traffic

To build the network structure for the computer communications we follow Ref. 20 and use data from the National Laboratory for Applied Network Research (NLANR) [25]. Here the network is constructed as follows: Over a period of 24 hours a number of servers associated to NLANR and physically spread over the USA, gathers information computer interconnections. For every connection established through a server the whole path, from the originating vertex to the requested destination, is added to the network graph. To be more specific the servers are using the Border Gateway Protocol (BGP) to relay connections over Internet's largest scale [26]. Vertices are computer networks, or "Autonomous Systems" in BGP nomenclature, interconnected by one or many BGP servers. An edge thus represents an established direct connection between two Autonomous Systems. The data we analyze, gathered on 28th December 1999, represent a network with $N = S = 2210$, $L = 4334$, and $\gamma \approx 0.221$.

### C. Erdös-Rényi model of random networks

In the Erdös-Rényi (ER) model [27], we start from $N$ vertices without edges. Subsequently, edges connecting two randomly chosen vertices are added until the total number of edges becomes $L$. It generates random networks with no particular structural bias: The only restriction in the model is that no multiple edges are allowed between two vertices. In this study, we choose the average degree $\langle k \rangle \equiv 2L/N$ as a control parameter in the ER model. The ER model graphs have a logarithmically increasing $\ell$, a Poisson-type degree distribution, and a clustering coefficient close to zero.

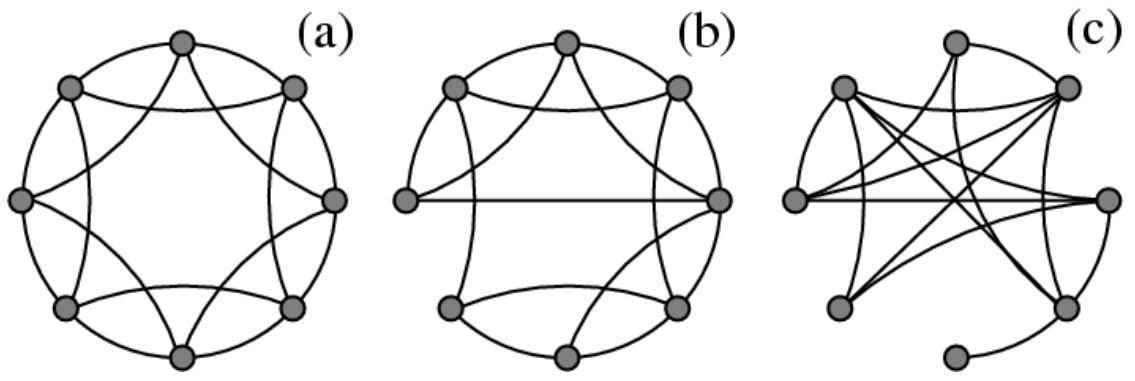

FIG. 3: The Watts-Strogatz (WS) model of small-world networks. The starting point is a regular one-dimensional lattice in (a) with the range $r = 2$ of the connections. Every edge is visited once and then with the rewiring probability $P$ is rewired to other vertex. The WS model can generate (a) the local regular network when $P = 0$, with the high clustering but with the large average geodesic length, and (c) the fully random network when $P = 1$, with low clustering but with very short geodesic length. In the intermediate region of $P$ depicted in (b), the WS model has *both* high clustering and the small-world behavior (more specifically, the average geodesic length $\ell \propto \log N$ for the network with the size $N$).

### D. Watts-Strogatz model of small-world networks

In the WS model [5] one starts by constructing a regular one-dimensional network with only local connections of range $r$. For example, $r = 2$ means that each vertex is connected to its two nearest neighbors and two next nearest neighbors (see Fig. 3(a)). Then each edge is visited once, and with the rewiring probability $P$ is detached at the opposite vertex and reconnected to a randomly chosen vertex forming a 'shortcut'. (See the illustration in Fig. 3.) For $P = 0$ the network is a regular local network, with high clustering, but without the small-world behavior: The average geodesic length in this case grows linearly with the network size. In the opposite limit of $P = 1$, where every edge has been rewired, the generated random graph has vanishing clustering, but shows a logarithmic behavior of the average geodesic length $\ell \propto \log N$. In an intermediate range of $P$ (typically $P \sim O(1/N)$), the network generated by the WS model displays *both* high clustering and small-world behavior—the commonly found characteristics of real social networks.

### E. Barabási-Albert model of scale-free networks

Apart from the average geodesic length and clustering, the degree distribution is a structural bias that has received much attention. Many (but not all) real networks are known to have a power-law distribution of degrees [3, 28], manifesting a scale-free nature of the network. The BA model of scale-free network [3, 20] is defined by the following ingredients:

- Initial condition: To start with the network consists of $m_0$ vertices and no edges.
- Growth: One vertex $v$ with $m$ edges is added every time step.
- Preferential attachment: An edge is added to an old vertex with the probability proportional to its degree. More precisely, the probability $P_u$ for a new vertex $v$ to be attached to $u$ is [33]:

$$P_u = \frac{k_u}{\sum_{w \in \mathcal{V}} k_w}. \tag{7}$$

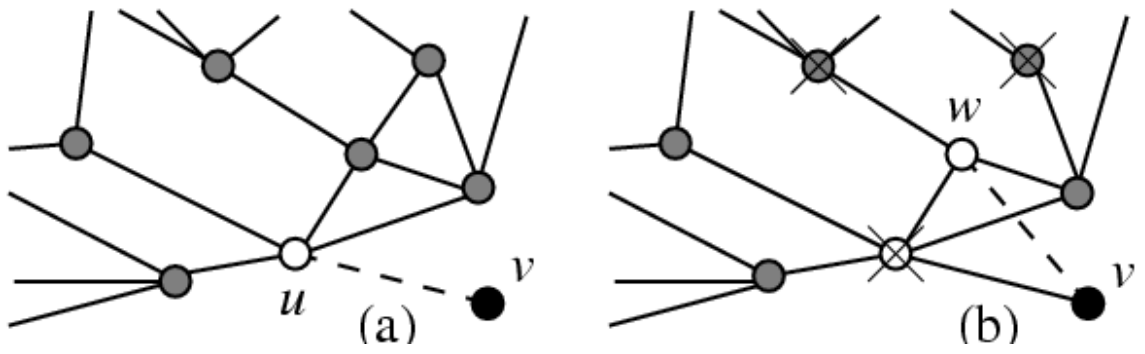

FIG. 4: The construction of the clustered scale-free (CSF) network in Sec. IV F. (a) In the preferential attachment step for the newly added vertex $v$ (denoted as filled black circle), the white vertex $u$ is chosen with the probability proportional to its degree (the dashed line represents the new edge). (b) In the triad formation step an additional edge (dashed line) is added to a randomly selected vertex $w$ in the neighborhood $\Gamma_u$ of the vertex $u$ chosen in the previous preferential attachment step in (a). The vertices marked by $\times$ are not allowed since they are not in $\Gamma_u$. Without the triad formation step, the CSF model reduces to the original BA model of scale-free networks.

The growth step is iterated $N - m_0$ times to construct the network with the size $N$, for each growth step the preferential attachment step is iterated $m$ times. The above described BA model has been shown to generate scale-free networks with the logarithmically increasing average geodesic length with the size $N$. However, the original BA model results in networks with low clustering.

### F. Clustered scale-free network model

In order to incorporate the high clustering of social networks one can modify the standard BA model by adding one additional step:

- Triad formation: If an edge between $v$ and $u$ was added in the previous step of preferential attachment, then add an edge from $v$ to a randomly chosen neighbor $w$ of $u$. This forms a triad, three vertices connected each other. If there is no available vertex to connect within $\Gamma_u$—do a preferential attachment step instead.

For every new vertex, after an additional preferential attachment step, the triad formation step is performed with a probability $P_t$ (and thus a preferential attachment with the probability $1 - P_t$). The average number of triad formation trials per added vertex $m_t \equiv (m-1)P_t$ is taken as a control parameter in this clustered scale-free (CSF) network model (see Fig. 4). The scale-free degree distribution of the original BA model is conserved in CSF model whose properties have been analyzed in detail in Ref. 4. In the limiting case of $m_t = 0$, the original BA network is constructed from the CSF model. The CSF model has been shown to exhibit the high clustering (furthermore the clustering coefficient is tunable by the control parameter $m_t$) while it still preserves the characteristics found in the BA model such as the logarithmically increasing average geodesic length and the scale-free degree distribution.

## V. CORRELATION BETWEEN DEGREE AND BETWEENNESS

For the six different networks described in Sec. IV, we seek the relation between the degree and the betweenness for vertices and edges. Both the degree and the betweenness, to some extent, measure how important the vertex (edge) is. The natural expectation is that the vertex (edge) with higher degree should also have higher betweenness. The calculation of the betweenness is based on the global information on paths connecting all pairs of vertices, while the degree, by definition, is the quantity which depends on only the local information. This implies that the identification of the relation between the degree and the betweenness can have a practical importance since one can approximately estimate the betweenness from the degree.

We first show in Fig. 5 scatter plots of the vertex betweenness $C_B(v)$ versus the vertex degree $k_v$. As expected, networks with the scale-free degree distributions, (a) the scientific collaboration network, (b) the computer network, (e) the BA network, and (f) the CSF network, show clear signs of correlation between the degree and the betweenness. As the scale-free network becomes more clustered (from the BA model to the CSF model), the correlation between $C_B(v)$ and $k_v$ becomes weaker, manifested by more scattered plots in (f) than (e). The ER and the WS models, (c) and (d), respectively, are characterized by the absence of vertices with very high degrees, which makes the correlation between $C_B(v)$ and $k_v$ rather difficult to observe especially in the region of high degrees. However, notable correlations are evident even for these networks with exponential cut-off in degree distributions.

For the study of the correlation between the edge degree $k_e$ and the edge betweenness $C_B(e)$, we try four different definitions of the edge degree in Eq. (6) from the assumption that the edge degree can be defined by only the degrees of vertices it connects. For all networks, ex-cept the scientific collaboration network, we find (at least some) correlation between $k_e$ and $C_B(e)$. This correlation is most evident with the definition in Eq. (6a): $k_e \equiv k_v k_w$, where $k_v$ and $k_w$ are degrees of vertices $v$ and $w$ which the edge $e$ connects. But the definition $k_e \equiv \min(k_v, k_w)$, Eq. (6c), also displays a high correlation between $k_e$ and $C_B(e)$. This suggests that the lower degree of the two vertices an edge connects, is more important for a high edge betweenness than the greater degree of the two vertices. In other words, this illustrates the quite natural situation that an edge does not necessarily become central just because it connects to one central vertex, rather it has to be a bridge between two central vertices.

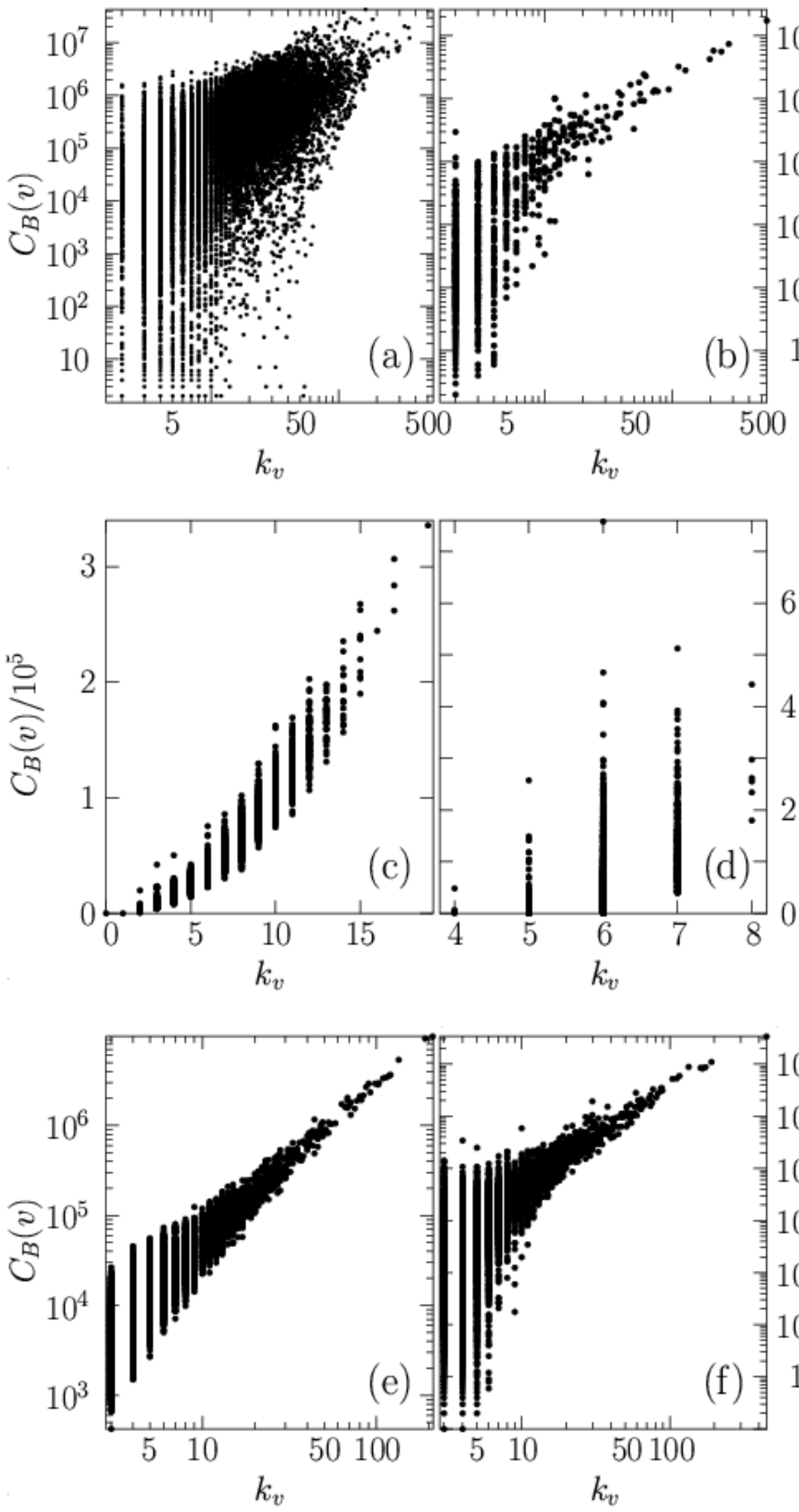

FIG. 5: Correlation between the vertex betweenness $C_B(v)$ and the vertex degree $k_v$ for (a) scientific collaboration network, (b) the computer network, (c) the ER network with the size $N = 10^4$, and the average degree $\langle k \rangle = 6$, (d) the WS network with $N = 10^4$, $r = 3$, and $P = 0.01$, (e) the BA network with $N = 10^4$, $m_0 = 5$, and $m = 3$, and (f) the CSF network with $N = 10^4$, $m_0 = 5$, $m = 3$, and $m_t = 1.8$; (see Sec. IV for details of networks). All are in log-log scales except for (c) and (d).

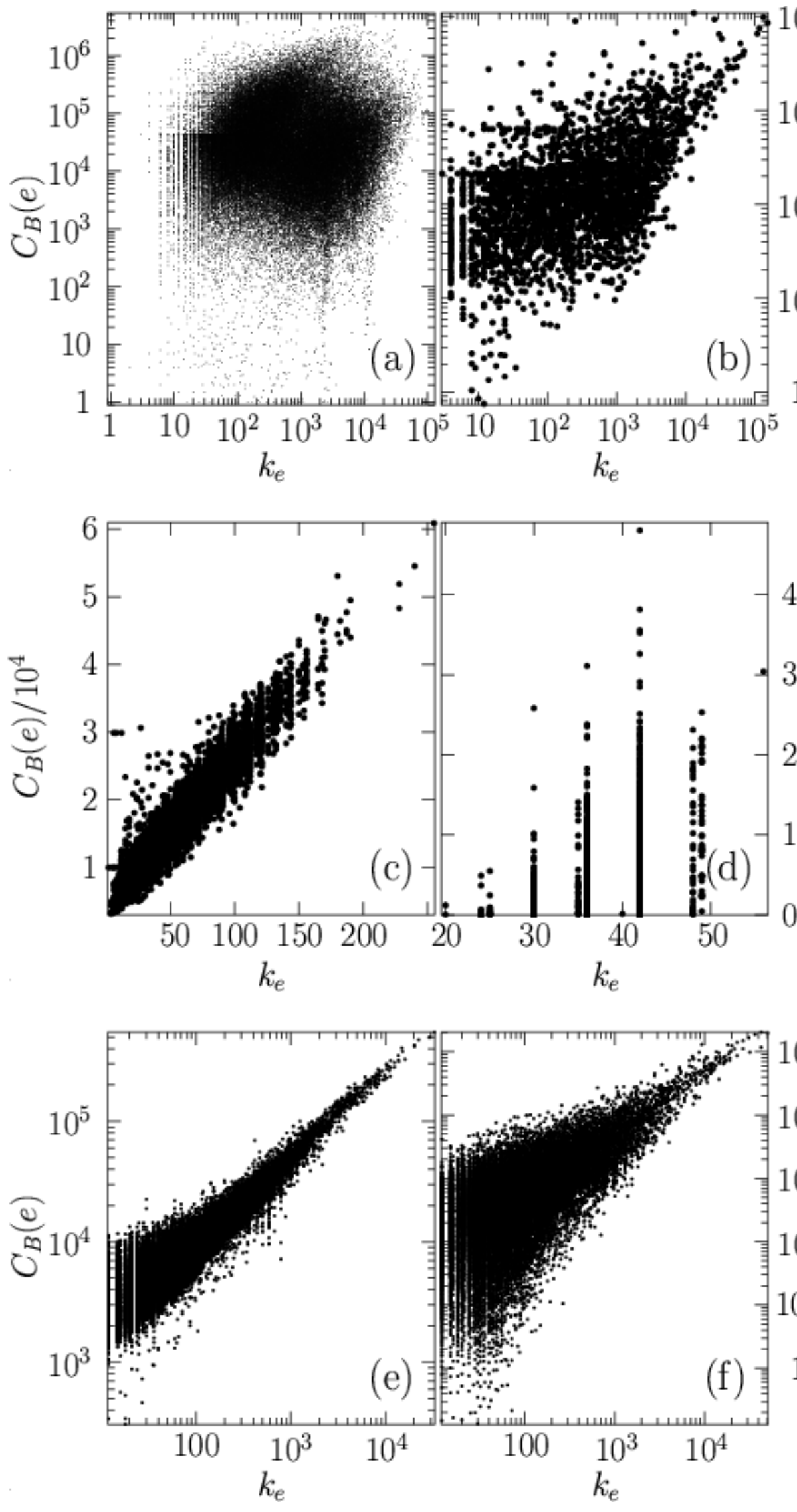

FIG. 6: Correlation between the edge betweenness $C_B(e)$ and the edge degree $k_e$ defined in Eq. (6a). The networks in this figure are identical to those in Fig. 5. Except (a) for the scientific collaboration network, all networks show clear correlation between the two quantities.

For the scientific collaboration network it turns out that none of the definitions of edge degree manifests the correlation clearly. Figure 6 shows the scatter plots for the edge degree and the edge betweenness, corresponding to the networks in Fig. 5. Especially, the similarity between the real network and the model network is evident between the computer network and the CSF network (compare (b) and (f) in Figs. 5 and 6), which suggests that the CSF model describes the network of computers better than the BA model. As far as the edge degree and betweenness are concerned, one can also argue that none of existing network model seems to de-scribe the scientific collaboration network properly, and the origin of the geometric difference in this network still remains to be studied.

## VI. VERTEX ATTACK

In this section, we study the attack vulnerability of six different complex networks described in Sec. IV by using the vertex attack strategies introduced in Sec. III, i.e., vertex removals using the information on the initial degree (ID), the initial betweenness (IB), the recalculated degree (RD), and the recalculated betweenness (RB). From the observation of the correlation between the vertex degree and the vertex betweenness in Sec. V, one expects that both betweenness-based and degree-based attack strategies should result in similar vulnerability behaviors. However, the detailed behaviors are found to show a variety of interesting differences among the networks. Figure 7 summarizes the results for the vertex attack vulnerability measured by the average inverse geodesic length $\ell^{-1}$ defined in Eq. (1) and the size $S$ of the giant component as functions of the number $N_{\rm rm}$ of removed vertices.

The two real-world networks, the scientific collaboration network and the computer network, show very distinct behavior: $\ell^{-1}$ and $S$ for the latter decay exponentially as shown in the inset in Fig. 7(b), whereas the former network (Fig. 7(a)) the decays of $\ell^{-1}$ and $S$ are almost linear. Another difference between the two real networks is that for the computer network the degree-based attack strategies (ID and RD) and the betweenness-based strategies (IB and RB) are almost equally harmful while the scientific collaboration network is more vulnerable to betweenness-based strategies (compare the insets of Fig. 7(a) and (b)). This behavior is somehow expected from the observation in Sec. V since in the region of high degrees (or in the early stage of attacks when most important vertices are removed) the computer network shows higher correlation between $k_v$ and $C_B(v)$ than the scientific collaboration network (compare Fig. 5(a) and (b)). In Fig. 7(a) for the scientific collaboration network, as the number $N_{\rm rm}$ of removed vertices is increased $\ell^{-1}$ for IB increases. This should not be interpreted as an enhancement of network functionality, but as an indication that the removed vertices are not the members of the largest connected subgraph (the giant component). In Fig. 7(a) the coherence between $\ell^{-1}$ and $S$ are also observable: Both show the jumps at the same places, when removal of one vertex results in the segmentation of the giant component. The removal procedures based on the initial network, ID and IB, are as expected not efficient for large $N_{\rm rm}$. At this point the network structures has changed so much compared to the initial network that the initially important vertices have lost their significance.

We next compare two model networks, the ER and the WS networks, which have exponential cut-off in the degree distributions. For the ER model in Fig. 7(c) the degree-based attack strategies prevail the betweenness-based ones. The strategies based on recalculated information are as expected more harmful than their counterparts based on the initial network. The two measured quantities $\ell^{-1}$ and $S$ rank the removal procedures in the same order. The WS small-world network model shows a completely different behavior than the ER model (see Fig. 7(d)). For small $N_{\rm rm}$ the RB procedure is the most harmful followed by the two degree based strategies. For the removal procedures based on the initial network the order is reversed—ID is more harmful than IB. These behaviors persist in the interval $0 \leqslant N_{\rm rm} \lesssim 2rNP$, where $2rNP$ is the number of endpoints of rewired edges, and other different behaviors emerge for larger $N_{\rm rm}$. This crossover behavior can be explained since when $N_{\rm rm}$ exceeds the number of the rewired edge endpoints the original WS model topology is lost. For larger $N_{\rm rm}$ the RB procedure retains its position as the most harmful procedure, although RD almost coincides with RB for $N_{\rm rm}/N \gtrsim 0.5$. The ID and IB procedures are already at an early stage quite harmless compared to RB as $\ell^{-1}$ starts to increase just as for the scientific collaboration network of Fig. 7(a). (A sign that mostly unimportant vertices are removed.) The most interesting and unexpected behavior occurs in an intermediate region around $N_{\rm rm}/N \approx 0.25$. Here the RD is the least harmful of all four procedures. It is also the only case where a recalculated information based procedure is less harmful than its counterpart based on the initial configuration. This clearly shows that choosing the vertex with highest degree is not an efficient way to destruct the WS small-world network. Recall that in the original network the degree is seldom far from $2r$.

The BA model (reviewed briefly in Sec. IV E) was in focus in the first study of the vulnerability of scale-free networks [20]. Scale-free networks are more sensitive to vertex removal than the ER and WS models. This is of course due to the large variation in the importance (measured both by degree and betweenness) of the vertices, i.e., there exist very important vertices which plays very important roles in the network functionality. In the ER and WS models the distribution of relevant measures of vertex importance, such as degree and betweenness, are restricted by the scale in the model. In the BA model shown in Fig. 7(e), the differences among the removal procedures are not significant in the early attack stage. However, as the removals proceed, the attack strategies harm network more in the order, RB > RD > ID > IB (the inequality RB > RD means that RB is more harmful than RD): As expected, strategies with recalculated degrees and betweennesses are more harmful. One interesting observation is the change of order between degree-based and betweenness-based strategies, i.e., (RB > RD versus ID > IB): This implies that the betweenness distribution changes more during the removal procedure than the degree distribution, which is natural since the betweenness depends on the global network structure, whereas the degree of a vertex is a quantity dependent on only its neighborhood.

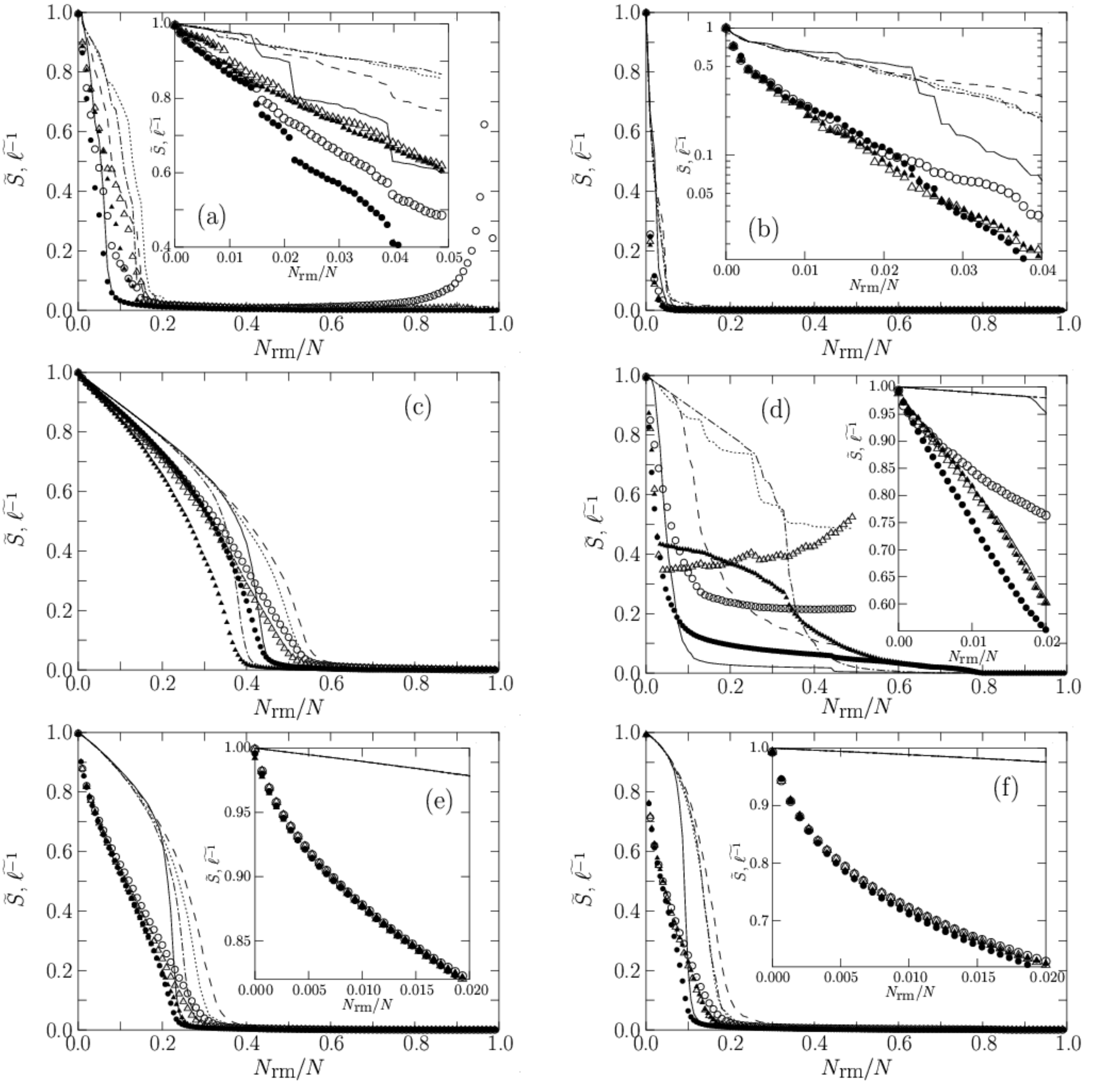

FIG. 7: Vertex attack vulnerability is measured by the average inverse geodesic length $\widetilde{\ell^{-1}}$ and the size of giant component $\widetilde{S}$ (see Sec. II), normalized by the values for initial networks, for (a) the scientific collaboration network, (b) the computer network, (c) the ER model with and $\langle k \rangle = 6$, (d) the WS model with $r = 3$, and $P = 0.01$, (e) the BA model with $m = 3$ and $m_0 = 5$, and (f) the CSF model with $m = 3$, $m_0 = 5$, and $m_t = 1.8$. $N = 1500$ for (c)-(f), see Sec. IV for details of networks. Four different attack strategies, each of which based on the initial degrees (ID), the initial betweenness (IB), the recalculated degree (RD), and the recalculated betweenness (RB), are used (see Sec. III). Empty symbols represent $\ell^{-1}$ obtained from attack strategies based on initial information, i.e, empty triangles (circles) for ID (IB), while filled symbols are for $\ell^{-1}$ from recalculated information, i.e., filled triangles (circles) for RD (RB). For model networks (c)-(f), the error bars estimated from the calculations for $\sim 40$ different network realizations are also plotted and they are smaller than the size of symbols. Lines represent $S$: Solid line represents RB, dash-dotted line represents RD, dotted line represents ID, and dashed line represents RB. Insets are magnifications in the early stages of attacks (note that inset in (b) is in the lin-log scale).

The CSF network model in Sec. IV F with tunable clustering is found to be even more attack vulnerable than the BA model as evident from the comparison of Fig. 7(e) and (f). This can be explained from the finding that for a fixed number of edges higher clustering makes the network less efficient. In other words, more geodesics go through the same important vertex when the network is highly clustered, making $\ell$ larger (or $\ell^{-1}$ smaller) [4], and at the same time making the network more vulnerable to the removal of an important vertex with many triads attached to it. Unlike the BA scale-free network model in Fig. 7(e), in the CSF network ID is not necessarily less harmful than RD for all $N_{\rm rm}$. In other words, the clustering makes the degree less important when assessing the vulnerability of a vertex. This can be explained by configurations such as the one in Fig. 8, which also causes the lower correlation between degree and betweenness mentioned above in Sec. V.

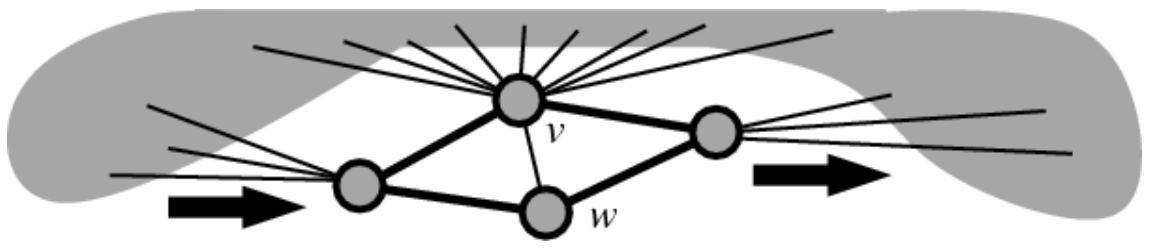

FIG. 8: Low degree vertex $w$ taking the load off a highly connected vertex $v$. It shows how high clustering can make connectivity less important for the routing of geodesics: $v$ with a high degree shares many geodesics with its low-degree neighbor $w$. The grey regions represents the part of the network that is connected to $v$. The arrows indicate that those geodesics can routed by either $v$ or $w$.

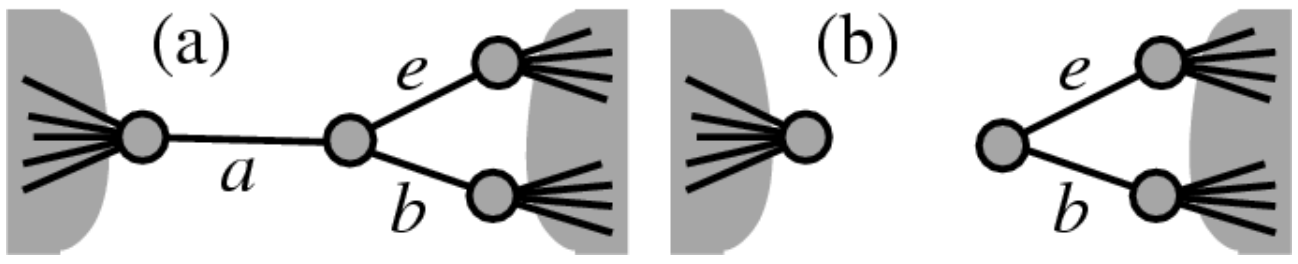

FIG. 9: Schematic diagram to show how an edge can lose or increase its betweenness. (a) The edge $a$ has a high betweenness as a bridge connecting the left and the right part of the network. (b) When the edge $a$ is removed, $e$ is only a part of geodesics in the right-hand side, and thus has lower betweenness than before the removal of $a$. On the other hand, if the edge $b$ had been removed, the betweenness of $e$ would have increased since all geodesics passing from the left to the right hand side would pass through $e$.

## VII. EDGE ATTACK

In the original study of the attack vulnerability of scale-free networks [20], only the vertex attacks have been considered, which may be interpreted as intruder-caused breakdowns of servers in computer networks, for instance. In this section, we study the vulnerability of complex networks subject to various types of edge attacks (see Sec. III for details of the edge attack strategies). For example, in computer networks, this can be interpreted as the malfunctioning or the loss of the communication cables. In the context of social networks, the attack on an edge can be interpreted as prohibition of contact between two individuals—a scenario that admittedly is somewhat artificial, but could possibly have practical implications to the prevention of spreading of sexually transmitted diseases [28]. As long as the network is not segmented into pieces, the average betweenness increases during edge attacks since the reduced number of edges should carry the same number of geodesics. However, this is not necessarily true for individual edges as illustrated in Fig. 9.

Figure 10 displays the results for the vulnerability to various edge attacks in six different complex networks. The symbols in Fig. 10 have the same meanings as in Fig. 7 in Sec. VI. Remember the different definitions of $\ell^{-1}$ for vertices and edges: When edges are removed the total number $N$ of vertices in the denominator in Eq. (1) does not change, making $\ell^{-1}$ a monotonously decreasing function with the number $L_{\rm rm}$ of removed edges.

The two real-world networks behave quite similarly under edge attacks as shown in Fig. 10(a) and (b), in contrast to vertex attacks where $\ell^{-1}$ decay linearly for the scientific collaboration network and exponentially for the computer network (see Sec. VI and Fig. 7(a) and (b)). For both real networks the RB procedure is the most destructive, followed by the IB procedure, in a broad range of $L_{\rm rm}$. The degree-based attack strategies, ID and RD, are found not as efficient as the betweenness-based ones, IB and RB, (in every case except large $L_{\rm rm}$ for the WS mode), which suggests that the edge betweenness is more suitable quantity than the edge degree to measure the importance of an edge. The correlation between the degree and the betweenness is stronger for vertices than edges (compare Fig. 5 and Fig. 6), which is related to the above finding that the edge degree fails to capture the importance of edges. We again find the similarity in the behavior of $\ell^{-1}$ and $S$ in Fig. 10(b): When a large part of the network becomes disconnected from the giant component, both $S$ and $\ell^{-1}$ show jumps simultaneously. On most occasions $S$ and $\ell^{-1}$ show the common features although each measures a distinct aspect of network performance: For example, the computer network in Fig. 10(b) at $L_{\rm rm}/L \approx 0.5$ exhibits the order of destructiveness RB > RD > IB > ID judging from $\ell^{-1}$, whereas $S$ gives the ranking RB > IB > RD > ID.

The two network models with the degree distributions characterized by exponential tails (the ER and the WS models), display again quite different vulnerability under edge attacks. For the ER model, judging from $\ell^{-1}$, the two attack strategies based on recalculated information (RD and RB) are the most harmful, and RB is clearly more harmful than all other three (see Fig. 10(c)). The RD curve for $S$ differs from the rest: $S$ is almost constant until $L_{\rm rm}$ reaches $L_{\rm rm}/L \approx 0.7$, where $S$ decreases very rapidly. The reason for this behavior is that the RD removal cuts edges between vertices that are highly connected, but the bridges—edges that, if removed, would disconnect the graph—are not necessarily linking vertices of high degree. Thus the structure emerging from repeated application of the RD strategy is characterized by a chain-like structure (low degrees of the vertices but relatively large connected subgraphs). This is also well illustrated by Fig. 2, where the maximal degree for the RD removal is $k_v = 2$ already at $L_{\rm rm} = 7$ (the same numbers for the other procedures are $L_{\rm rm} = 9$ for ID and RD, and $L_{\rm rm} = 10$ for RB). The graph is also connected longer for the ID and RD strategies than for the betweenness-based strategies.

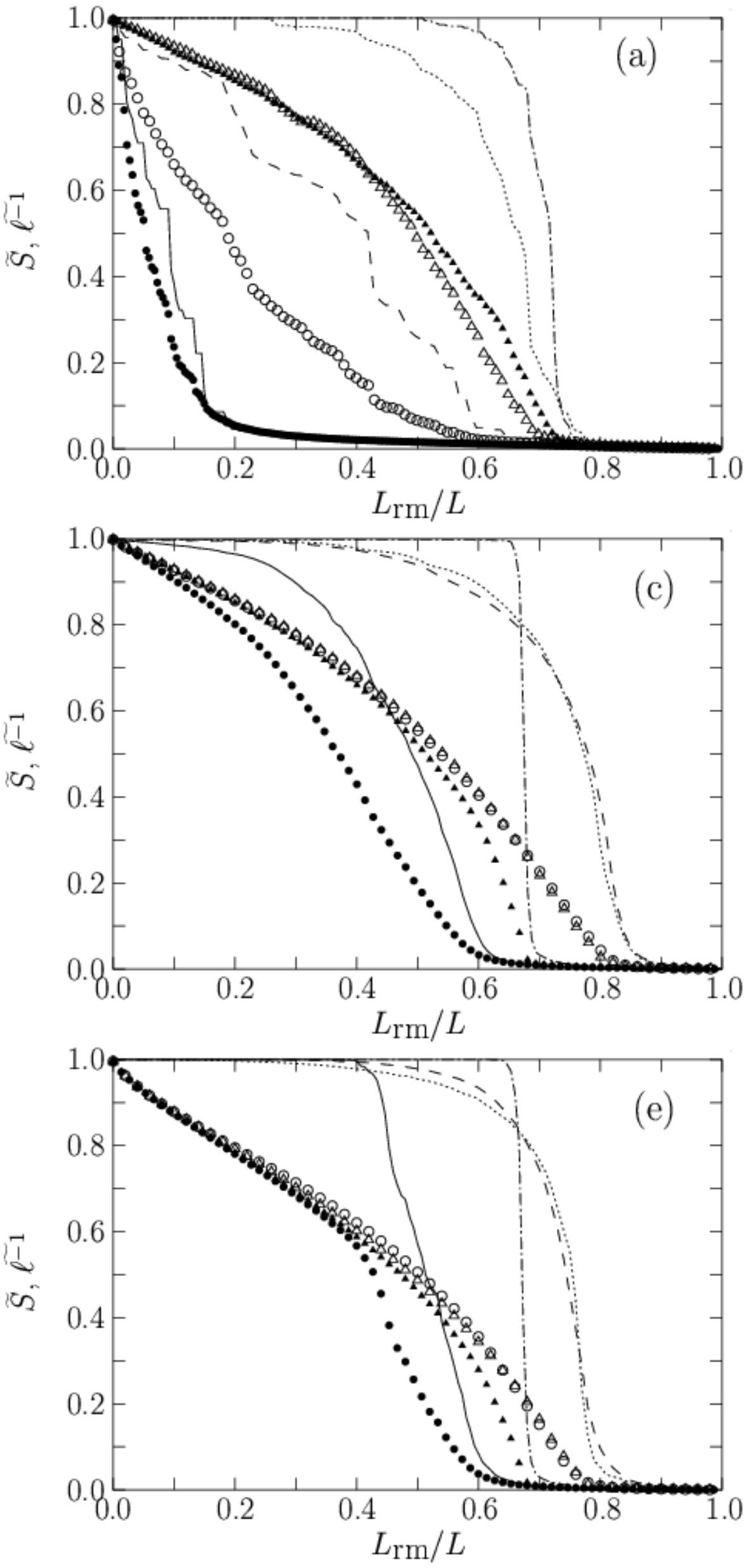

FIG. 10: Edge attack vulnerability for the same networks as in Fig. 7 subject to four different edge attack strategies (see Sec. III). $L_{\rm rm}$ is the number of removed edges and the other symbols and notations are the same as in Fig. 7. For the degree-based edge attack strategies, the definition of the edge degree in Eq. (6a) has been used.

Just as for the vertex attacks the WS network model again shows a very different behavior (see Fig. 10(d)). For the three procedures ID, IB, and RB, the inverse geodesic length $\ell^{-1}$ decays very rapidly for small $L_{\rm rm}/L \lesssim 0.07$. In the WS model the rewired edges are carrying a large portion of geodesics [2]. The vertices which are endpoints of rewired edges have higher vertex degree ($\langle k \rangle + 1$) and accordingly the edges connected to those endpoint vertices have also higher edge degree ($\langle k \rangle(\langle k \rangle + 1)$), becoming early targets of edge attacks. Since there exist $PL$ rewired edges, the number of edges with higher edge degree in initial network is of course $(\langle k \rangle + 1)PL$. Consequently, for $0 \leqslant L_{\rm rm} \lesssim (\langle k \rangle + 1)PL$ ($= 0.07L$ in our case) the network is very vulnerable to edge attack—in fact, for these three procedures (ID, IB, and RB) the WS network is more vulnerable than any other networks we study. On the other hand, for $(\langle k \rangle + 1)PL \lesssim L_{\rm rm}$ the network topology has lost most resemblance to the original network, and in this region the decay of $\ell^{-1}$ is far less rapid than for small $L_{\rm rm}$. The behavior of the RD removal is strikingly different: For $L_{\rm rm}/L \lesssim 0.01$ the decay of $\ell^{-1}$ is as rapid as the other three procedures, but for $0.01 \lesssim L_{\rm rm}/L \lesssim 0.07$ the decay is much slower. That this behavior is relevant in the large-$N$ limit is shown in Fig. 11. If an edge $e = (v, w)$ is rewired and there are not other rewired edges attaching to $v$ or $w$, the edge degree of $e$ is $k_e = \langle k\rangle(\langle k\rangle + 1)$, which is larger than the average $k_e = \langle k\rangle^2$. The RD removal will thus first attack the rewired edges, just as the ID removal (and probably IB and RB as well), but for $L_{\rm rm} \gtrsim 0.01$ the RD removal picks less vulnerable edges than the ID strategy. The reason for this behavior can be discussed in the context of Fig. 12, where the edge $(w, u)$ rewired to $(w, v)$ is the only rewired one: The ID removal will remove edges in the neighborhood $\Gamma_v$ (which has $k_e = \langle k\rangle(\langle k\rangle + 1)$) including the rewired edge first, followed by edges outside $\Gamma_u \cup \Gamma_v$ (with $k_e = \langle k\rangle^2$) and at last the edges in $\Gamma_u$ (with $k_e = \langle k\rangle(\langle k\rangle - 1)$). The RD strategy will start by removing one edge in $\Gamma_v$, not necessarily the rewired one. Say $(v', v)$ is the first removed edge (where $v'$ could be equal to $w$), second edge to remove can be anyone outside $\Gamma_u \cup \Gamma_{v'}$. The RD strategy strives to keep the degree uniform, which leads to a twofold disadvantage compared to the ID strategy: Firstly, after $\langle k\rangle + 1$ removed edges the rewired edge is removed with certainty by the ID removal, but only with a probability $\sim 1/(\langle k\rangle + 1)$ by the RD removal. Secondly, that the RD strategy keeps the degree uniform is also preventing its efficiency—compare the dashed-dotted (RD) line with the dashed (ID) line of Fig. 10(d). The mechanism of this can also be seen from Figs. 12(b) and (c): With the ID removal all edges in $\Gamma_v$ are deleted when $L_{\rm rm} = \langle k\rangle + 1$ (see Fig. 12(b)), but for the RD removal this is not the case (see Fig. 12(c)). The part of the graph shown in Fig. 12(b) can be disconnected by removing three edges, in Fig. 12(c) the same number is four, but all of these have a lower degree than the edges not in a neighborhood of an removed edge and will thus be removed later.

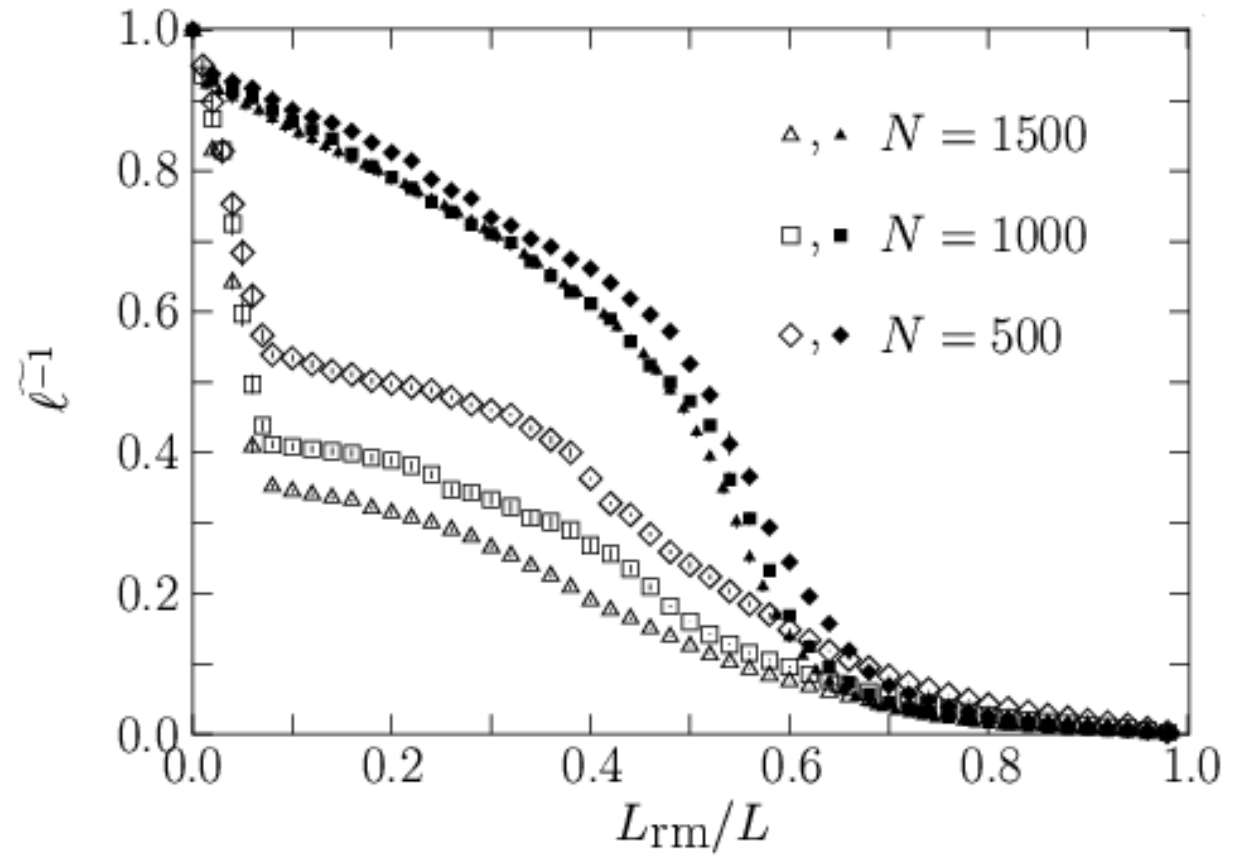

FIG. 11: Average inverse geodesic length $\widetilde{\ell^{-1}}$ for edge attack on the WS small-world network model with sizes $N = 500$, 1000, and 1500 subject to the RD removal strategy (filled symbols) and IB removal strategy (empty symbols). As $N$ becomes larger, $\widetilde{\ell^{-1}}$ is shown to saturate with a remarkable difference between RD and ID.

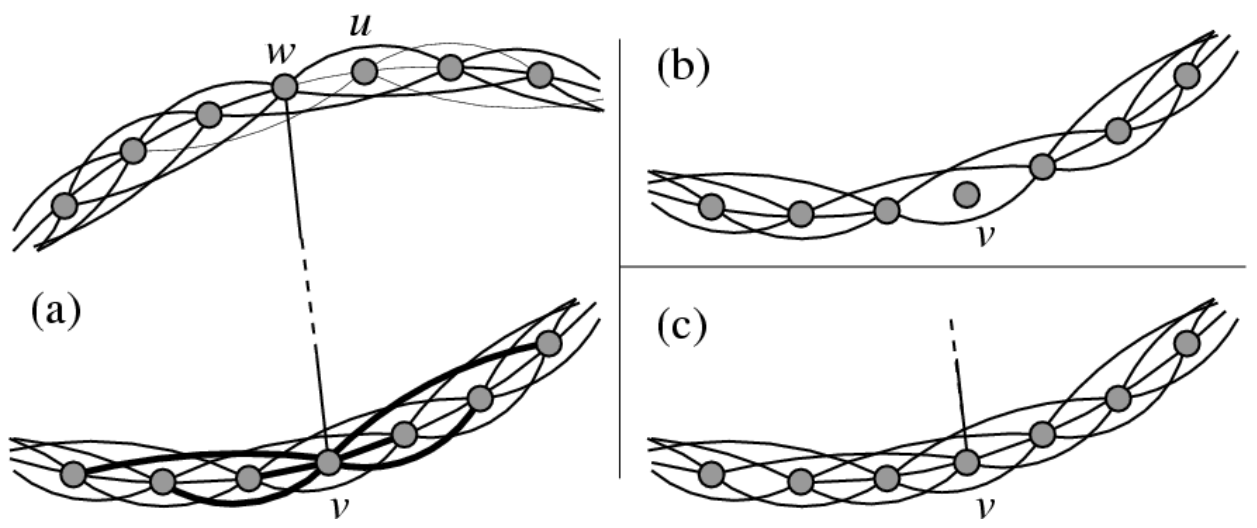

FIG. 12: RD vs ID strategies on a WS model network with $\langle k\rangle = 6$. the original configuration is shown in (a). The vertically oriented edge in (a) starting at $w$ is rewired from $u$ to $v$. In (a), the thicker lines connected to $v$ represent edges with higher edge degree, and the thinner lines connected to $u$ represent edges with lower degree. Typical configurations in the lower part of (a) after $\langle k\rangle + 1$ (in this case seven) removed edges are shown in (b) for ID removal and (c) for RD removal, leading to the prediction that RD is much less harmful than ID as confirmed in Fig. 10(d).

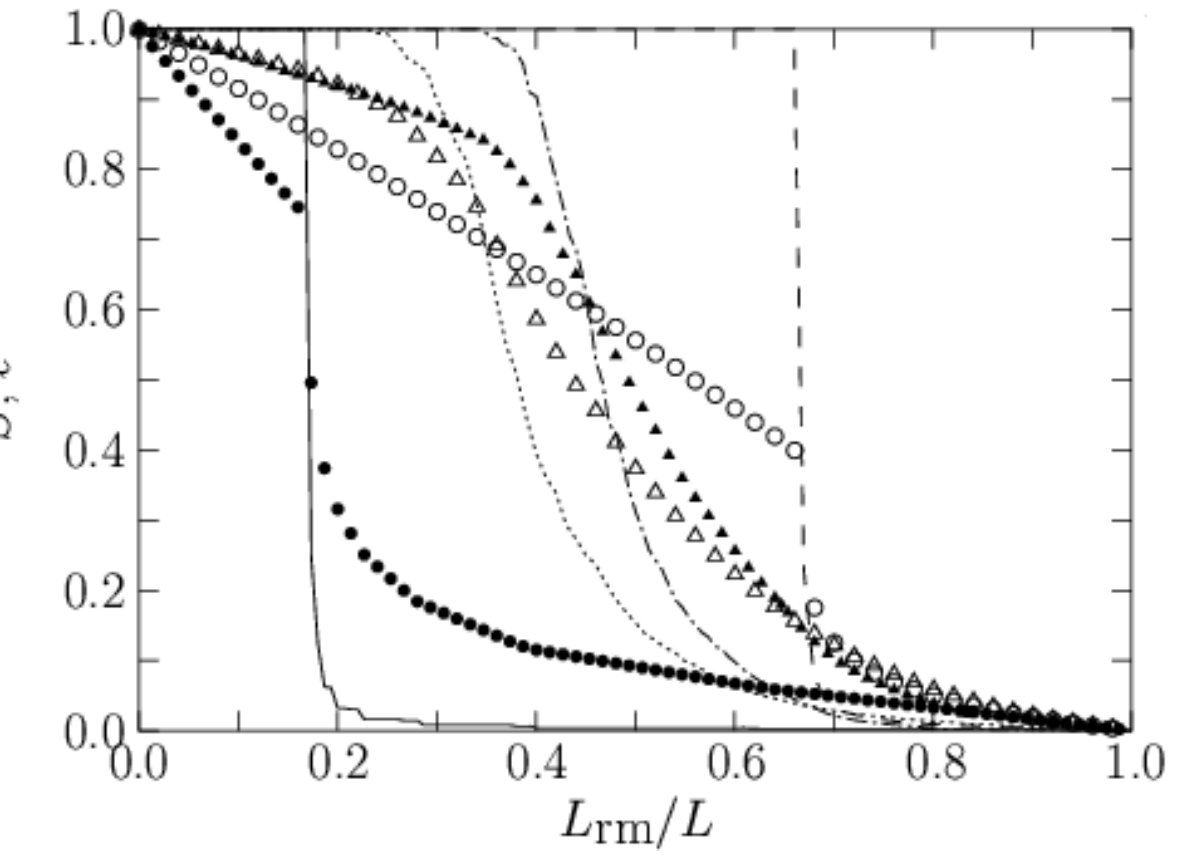

FIG. 13: Edge attack on a regular one-dimensional network with connection range $r = 3$ and $N = 1500$. Symbols are the same as in Figs. 7 and 10.

To illustrate the effect of the rewiring in the construction of the WS network we briefly discuss edge attacks on a regular network. For IB removal we note that the betweenness of the initial regular network is increasing rapidly with the range of the edges. This means that at $L_{\rm rm} = (r - 1)L/r$ the graph will be just a connected ring of $k_v = 2$ vertices. After this it will disintegrate rapidly. As shown in Fig. 13 for the regular network with $r = 3$, this behavior is seen around $L_{\rm rm} = 2L/3$. Since the initial degree of the regular network is uniform, the edges are removed in a completely random order for the ID procedure. This makes the graph lose its connectedness at around $L_{\rm rm} \approx L/4$. When a regular network is attacked by the RD procedure, the edge and vertex degree will be kept as uniform as possible. The network will remain connected until edges of degree $k_e = (\langle k\rangle/2 + 1)\langle k\rangle/2$ will be removed (since the network cannot be disconnected unless four vertices of degree $\langle k\rangle/2$ exist). This happens with $(1 + \langle k\rangle/2)N/2 - 4$ edges left in the system, so in our system $S = 1$ at least until $L_{\rm rm}/L \approx 0.33$. In Fig. 13 this happens roughly at $L_{\rm rm}/L \approx 0.36$, which is later than for the ID removal (where $S = 1$ up to $L_{\rm rm}/L \approx 0.24$). This effect is also present for the WS-model networks (mentioned as our second point above). Comparing Fig. 10(d) and Fig. 13 one notices that the area of rapid decay of $\ell^{-1}$ ($0 \leqslant L_{\rm rm}/L \lesssim P$ for IB, RB, and RD; and $0 \leqslant L_{\rm rm}/L \lesssim (\langle k\rangle + 1)P$ for ID) is lacking for the regular network, confirming that the rewired edges are responsible for this strong vulnerability. Another observation is that the cusps arising from the regularity of the network in the $\widetilde{\ell^{-1}}$ curves for IB and RB removal in Fig. 13 are gone in Fig. 10(d). This is of course expected since the number of rewired edges is different for different disorder realizations.

For the case of vertex attack, the two network models with scale-free degree distributions display a rather similar behavior. For edge attack there are larger differences. As seen in Fig. 10(e), for the BA scale-free network model the recalculated attack strategies are the most harmful. The differences between the four methods are not very large for $0 \leqslant L_{\rm rm}/L \lesssim 0.4$. This suggests that the characteristic topology of the BA scale-free network model is retained for this rather broad region. Just like for the ER model $S$ for the RD procedure decreases very rapidly at $L_{\rm rm}/L \approx 0.7$, crossing ID and IB curves. For $L_{\rm rm}/L \lesssim 0.4$ $S$ for ID and IB are actually even below the RB curve. Similarly to the vertex attack, the CSF model proves to be more vulnerable than the BA model (see Fig. 10(d)). Just like the other networks with high clustering (the scientific collaboration network and the WS model), the RD procedure is the least harmful for the CSF network. For the three unclustered networks, to which the BA mode belongs, the RD strategy is not the worst. Once again this can be explained by the low correlation between degree and betweenness for clustered networks (see Sec. V and Fig. 8) The most drastic difference between the scale-free network models is the curves for the RB procedure, which is the most harmful selection procedure. Here $\ell^{-1}$ for the CSF network shows the same positive convexity as the scientific collaboration network of Fig. 10(b). The $\ell^{-1}$ curve for the IB procedure has, however, a negative convexity in Fig. 10(d), as opposed to Fig. 10(b). This could be guessed from that the maximum betweenness is much higher for the CSF model (Fig. 6(f)) than the BA model (Fig. 6(e)). Our conclusion is that, although the CSF model with tunable clustering shows the closest resemblance to the scientific collaboration network, and this is presumably thanks to the high clustering and scale-free degree distribution, there are structures governing real-world networks, which are yet to be quantified.

In summary of the different networks under various attack strategies, measured quantities are shown in Table VII. The values of $\widetilde{\ell^{-1}}$ and $\widetilde{S}$ are shown after 1% of the vertices (or edges) are removed, i.e., $N_{\rm rm}/N = L_{\rm rm}/L = 0.01$. This is chosen to be small enough to keep the most original network structure, but large enough to display the changes introduced by various types of attacks. The first criterion obviously fails for the computer network, but the value 1% is used anyway for the sake of comparison with other networks.

## VIII. SUMMARY AND CONCLUSIONS

We have studied the correlation between degree and betweenness for both vertices and edges (see Fig. 5). For vertex degree and vertex betweenness the correlation is very strong for the ER model and the BA scale-free network, and is also evident in the WS small-world network model and the scale-free network model with tunable clustering. Of the real-world networks the Internet traffic network shows a strong correlation, whereas the scientific collaboration network has weaker correlation. For edges we define an edge degree as the product of the degrees of the linked vertices. The similar scatter plots show weaker correlation, a result of the lack of natural generalization of the degree concept from vertices to edges.

Computer network shows a unique behavior when subject to vertex attack—the average inverse geodesic length $\ell^{-1}$ clearly shows an exponentially decay in the early stages of attacks. Scientific collaboration network, in contrast, shows a linear decay for the same quantity. For edge attack on real-world networks the recalculated betweenness (RB) procedure is the most efficient. The difference between the attack strategies based on the initial information and the recalculated information shows that the change of network structure during the removal process is substantial. This must be taken into consideration if one wants to give efficient protection to a network.

None of the network models shows a behavior very similar to the real-world networks: Even the clustered scale-free network model with both high clustering and the scale-free degree distribution, which are two important characteristics in real-world networks, fails to describe successfully the scientific collaboration network. This clearly suggests that there are other structures contributing to the network behavior during vertex attack, and conclusions from model networks should be cautiously generalized to real-world situations. However, it should be emphasized that the CSF model under edge attacks with the RB strategy shows a behavior similar to the highly clustered scientific collaboration network (compare Fig. 10(a) and Fig. 10(f)), whereas the BA model with very low clustering in Fig. 10(e) shows clearly different behavior under the same RB edge attack.

The ER model, that lacks structural bias, is the most robust of the tested networks. This supports the intuitive idea that building a server-less network would be very robust to attack. Even if the network connections would be fixed in a random pattern this would lead to a tremendous increase of attack-robustness of the net-work (as the ER model shows). Wireless and server-less networks, so called "ad-hoc networks", are well studied from a theoretical viewpoint [29]. Most literature on network security concerns software protection and prevention of loopholes [30], rather than the network topology. This is of course natural since it is the primary defense against computer network attack (along with locking the computer room). But as a background protection, an attack-robust network topology can be useful; and thus we believe that the robustness of server-less networks should encourage further research.

| | | | | $\widetilde{\ell^{-1}}[\widetilde{S}]$ for vertex attack | | | | $\widetilde{\ell^{-1}}[\widetilde{S}]$ for edge attack | | | |
|---|---|---|---|---|---|---|---|---|---|---|---|
| Network | $N$ | $\langle k \rangle$ | $\gamma$ | ID | IB | RD | RB | ID | IB | RD | RB |
| SC | 2010 | 6.6 | 0.57 | 0.90 [0.96] | 0.88 [0.98] | 0.89 [0.96] | 0.86 [0.98] | 0.98 [1.00] | 0.92 [0.98] | 0.95 [1.00] | 0.88 [0.95] |
| CN | 2122 | 4.1 | 0.22 | 0.20 [0.62] | 0.25 [0.63] | 0.20 [0.62] | 0.24 [0.66] | 0.83 [1.00] | 0.74 [0.93] | 0.82 [1.00] | 0.51 [0.68] |
| ER | 1500 | 6.0 | 0.0040 | 0.98 [0.99] | 0.98 [0.99] | 0.97 [0.99] | 0.98 [0.99] | 0.99 [0.99] | 0.99 [0.99] | 0.99 [1.00] | 0.99 [1.00] |
| WS | 1500 | 6.0 | 0.58 | 0.82 [0.99] | 0.85 [0.99] | 0.82 [0.99] | 0.75 [0.99] | 0.89 [1.00] | 0.65 [1.00] | 0.92 [1.00] | 0.54 [1.00] |
| BA | 1500 | 6.0 | 0.015 | 0.88 [0.99] | 0.88 [0.99] | 0.88 [0.99] | 0.88 [0.99] | 0.98 [1.00] | 0.98 [1.00] | 0.98 [1.00] | 0.98 [1.00] |
| CSF | 1500 | 6.0 | 0.54 | 0.72 [0.99] | 0.71 [0.99] | 0.72 [0.99] | 0.70 [0.99] | 0.93 [1.00] | 0.94 [1.00] | 0.93 [1.00] | 0.93 [1.00] |

TABLE I: The normalized average geodesic length $\widetilde{\ell^{-1}}$ and the normalized size of the largest connected component $\widetilde{S}$ computed after 1% of the vertices (edges) are removed. The normalization is made to satisfy $\widetilde{\ell^{-1}} = \widetilde{S} = 1$ for the original networks. SC, CN, ER, WS, BA, and CSF denote the scientific collaboration network, the computer network, the random network model by Erdös and Rényi, the Watts-Strogatz model of the small-world, the Barabási-Albert model of the scale-free network, and the clustered scale-free network model in Ref. 4 (see Sec. IV for details of networks).

Lacking in this study, and an interesting area for future studies, is an extensive scaling analysis to establish the borders in parameter space for the different responses to the attack procedures. (We study one case with this method when we, from Fig. 11, conclude that the ID strategy is more efficient than the RD strategy in the $N \to \infty$ limit.)

### Acknowledgements

The authors would like to thank Prof. M. E. J. Newman for providing the scientific collaboration data, and Dr. M. Ozana for discussions. P. H. and B. J. K. acknowledge support by the Swedish Natural Research Council through Contracts Nos. F 5102-659/2001 and E 5106-1643/1999. C. N. Y. and S. K. H. acknowledge the support from KOSEF.